\documentclass{aa}

\usepackage[varg]{txfonts}
\usepackage{natbib}

\usepackage{graphicx}
\usepackage{longtable}

\begin{document}
\title{On the incidence of eclipsing Am binary systems in the SuperWASP survey}
\author
{B.~Smalley\inst{1},
J.~Southworth\inst{1},
O.I.~Pintado \inst{2}\thanks{Visiting Astronomer, Complejo Astron\'{o}mico El Leoncito operated under
agreement between the Consejo Nacional de Investigaciones Cient\'{i}ficas y
T\'{e}cnicas de la Rep\'{u}blica Argentina and the National Universities of La
Plata, C\'{o}rdoba and San Juan.},
M.~Gillon\inst{3},
D.~L.~Holdsworth\inst{1},
D.~R.~Anderson\inst{1},
S.~C.~C.~Barros\inst{4},
A.~Collier Cameron\inst{5},
L.~Delrez\inst{3},
F.~Faedi\inst{6},
C.~A.~Haswell\inst{7},
C.~Hellier\inst{1},
K.~Horne\inst{5},
E.~Jehin\inst{3},
P.~F.~L.~Maxted\inst{1},
A.~J.~Norton\inst{7},
D.~Pollacco\inst{6},
I.~Skillen\inst{8},
A.~M.~S.~Smith\inst{9},
R.~G.~West\inst{6},
P.~J.~Wheatley\inst{6}}

\authorrunning{B. Smalley et al.}
\titlerunning{Eclipsing Am binary systems}
\institute{
Astrophysics Group, Keele University, Staffordshire, ST5 5BG, United Kingdom
\and
Instituto Superior de Correlaci\'on Geol\'ogica-CONICET, 4000 Tucum\'an, Argentina 
\and
Institut d'Astrophysique et de G\'{e}ophysique, Universit\'{e} de Li\`{e}ge,
All\'{e}e du 6 ao\^{u}t 17, Sart Tilman, Li\`{e}ge 1, Belgium
\and
Aix-Marseille Universit\'{e}, CNRS, LAM (Laboratoire d'Astrophysique de Marseille) UMR 7326, 13388 Marseille, France
\and
SUPA, School of Physics \& Astronomy, University of St. Andrews,
North Haugh, Fife, KY16 9SS, UK
\and
Department of Physics, University of Warwick, Coventry, CV4 7AL, UK
\and
Department of Physical Sciences, The Open University, Walton Hall, Milton
Keynes, MK7 6AA, UK
\and
Isaac Newton Group of Telescopes, Apartado de Correos 321, 38700 Santa Cruz de
la Palma, Tenerife, Spain
\and
N. Copernicus Astronomical Centre, Polish Academy of Sciences, Bartycka 18, 00-716 Warsaw, Poland
}

\date{Received date / accepted date}

\abstract{The results of a search for eclipsing Am star binaries using
photometry from the SuperWASP survey are presented. The light curves of 1742 Am
stars fainter than $V$ = 8.0 were analysed for the presences of eclipses. A
total of 70 stars were found to exhibit eclipses, with 66 having sufficient
observations to enable orbital periods to be determined and 28 of which are
newly identified eclipsing systems. Also presented are spectroscopic orbits for
5 of the systems. The number of systems and the period distribution is found to
be consistent with that identified in previous radial velocity surveys of
`classical' Am stars.}


\maketitle

\section{Introduction}
\label{introduction}

Amongst the A and F stars there exists a subclass of peculiar stars called the
metallic-lined (Am) stars, in which the \ion{Ca}{II}~K line is considerably
weaker than would be expected from the average metallic line type
\citep{1940ApJ....92..256T,1948ApJ...107..107R}. These stars exhibit an apparent
underabundance of calcium and scandium, overabundances of iron-group elements,
and extreme enhancements of rare-earth elements \citep{1970PASP...82..781C}. In
contrast to normal A-type stars, the Am stars are slowly rotating
\citep{1983aspp.book.....W} with maximum $v \sin i$ values of $\sim$100\,km\,s$^{-1}$
\citep{1973ApJ...182..809A}. The abundance anomalies are thought to be due to
radiative diffusion of elements within the stable atmospheres of these
relatively slowly rotating stars
\citep{1970ApJ...160..641M,1980AJ.....85..589M,1983ApJ...269..239M}.

Early spectral studies of Am stars hinted at a high fraction of spectroscopic
binaries \citep{1948ApJ...107..107R}, while the systematic study by
\cite{1961ApJS....6...37A} led to the conclusion that all Am stars are members
of spectroscopic binaries. Hence, it was assumed that the slow rotation of Am
stars, required for radiative diffusion to occur, was the result of the
reduction of rotational velocities due to tidal interaction. 
While there were spectroscopic orbits for many Am stars \citep[e.g.][]{2004A&A...424..727P}, only a
handful were known to be eclipsing \citep[e.g.][]{1980ARA&A..18..115P,1991A&ARv...3...91A}.
It was this that led \cite{1990clst.book.....J} to conclude that:
\begin{quote}
\emph{A curious fact is that among the many Am stars known (all of which are binaries)
there should be many eclipsing binaries, but surprisingly very few cases are
known.}
\end{quote}

Comprehensive spectroscopic radial velocity studies of carefully selected Am
stars, have found a binary fraction of nearer 60--70\%
\citep{1985ApJS...59..229A,2007MNRAS.380.1064C}. The period distribution shows
that the majority of systems have periods $\la$\,50\,days, consistent with the 
slow rotation being due to tidal synchronisation or pseudo-synchronization
\citep{1996A&A...313..523B,1997A&A...326..655B}. There are, nonetheless, systems
with longer periods, suggesting that these Am stars were formed with low initial
rotation velocities \citep{2004ASPC..318..297N}. The key to the Am phenomenon
appears to be slow rotation and not binarity per se.

The recent \citet{2009A&A...498..961R} catalogue lists only 61 Am stars as
eclipsing or possibly eclipsing. This represents only 1.4\% of the Am stars in
the catalogue. Given that a large fraction of Am stars are supposed to be in
binary systems, this percentage does appear rather low. For example, in a binary
with a period of $\sim$5\,days (typical of many Am spectroscopic binaries), there
is a $\sim$10\% probability that the system should be eclipsing. Hence, there is
a perhaps somewhat na\"{i}ve expectation that there ought to be more eclipsing Am stars. It is
this which led us to investigate the number of eclipsing systems that can be
found using light curves obtained from the SuperWASP exoplanet transit survey.

\section{SuperWASP observations}

The WASP project is surveying the sky for transiting extrasolar planets
\citep{2006PASP..118.1407P} using two robotic telescopes, one at the
Observatorio del Roque de los Muchachos on the island of La Palma in the Canary
Islands, and the other at the Sutherland Station, South African Astronomical
Observatory (SAAO). Both telescopes consist of an array of eight 200-mm, f/1.8
Canon telephoto lenses and Andor CCDs, giving a field of view of $7.8\degr
\times 7.8\degr$ and pixel size of around 14\,\arcsec. The observing strategy is
such that each field is observed with a typical cadence of the order of 10\,min.
WASP provides good quality photometry with a precision exceeding 1\% per
observation in the approximate magnitude range $9 \le V \le 12$.

The SuperWASP data reduction pipeline is described in detail in
\cite{2006PASP..118.1407P}. The aperture-extracted photometry from each camera
on each night are corrected for primary and secondary atmospheric
extinction, instrumental colour response and system zero-point using a
network of stars with colours defined in the Tycho-2 catalogue.
Even though the WASP bandpass extends further into the red, the
resultant pseudo-$V$ magnitudes are comparable to Tycho $V$ magnitudes.
Additional systematic errors affecting all the stars are identified and removed
using the SysRem algorithm of \cite{2005MNRAS.356.1466T}.

We have selected Am stars from the \citet{2009A&A...498..961R}
catalogue\footnote{We use the prefix Renson to refer to entries in the
\cite{2009A&A...498..961R} catalogue} for
which we have data in the WASP archive and when individual light curves have at
least 1000 data points (i.e. for a single camera and during a single season). 
In addition, we rejected any stars with magnitudes brighter than $V=8.0$, in
order to avoid the most significant effects of saturation in the WASP images. A
total of 1742 stars were selected for light curve analysis, which is 55\% of the
Am stars of $V=8.0$ or fainter in \citet{2009A&A...498..961R} and 40\% of all
the Am stars in the catalogue.

\section{Light curve analysis}

The light curves of the target stars were analysed using the WASP Project's
\textsc{hunter} program \citep{2006MNRAS.373..799C}, which is an adaptation of
the Box Least Squares algorithm of \cite{2002A&A...391..369K}. The algorithm
computes $\chi^2$ values of transit model light curves using a box-shaped model
that is slid over the observed light curve. The period-searching range was from
1 to 50 days. Each individual light curve was then folded on the periods of the
5 most significant $\chi^2$ values and visually inspected for the presence of
eclipses.

From the survey of 1742 Am stars, fainter than $V$ = 8.0, 70 eclipsing systems
were found, of which 28 are previously unreported detections
(Table\,\ref{table:Am Binaries}) and 4 are suspected eclipsing systems
(Table\,\ref{table:Am Binaries No Period}), but with too few eclipses to confirm
their eclipsing status or to determine an orbital period. This brings the total
number of known eclipsing Am stars to around 100. Around 4\% of the Am stars in
our sample have been found to exhibit eclipses.

\begin{table*}
\caption{Eclipsing Am binaries. Columns 5 and 6 give the depths, in {mmag.}, of primary
(Min\,\textsc{i}) and secondary (Min\,\textsc{ii}) minima. Column 7 gives the phase of
secondary minimum ($\phi_{\mbox{\textsc{ii}}}$) if different from 0.50. Column 8 gives the binary classification. A
dagger (\dag) indicates that there is a possible $P\times2$ uncertainty from the
WASP light curve. For previously known systems Column 10 gives the GCVS
designation where available, otherwise either an ASAS designation
\citep{2002AcA....52..397P} or a literature reference.}
\label{table:Am Binaries}
\scriptsize
\begin{tabular}{llllllllll} \hline\hline

Renson&Name&Spec. Type&$P_{\rm orb}$ (d)&Min I&Min II&$\phi_{\mbox{\textsc{ii}}}$&Class&Dilution&Known System\\
\hline
 3330 & HD\,12950       & A4mA8                 & 2.39831 & 0.029 & 0.018 & ...  & Ell              & 0.55 & \\
 3590 & HD\,14111       & A0mF2                 & 1.63078 & 0.042 & 0.009 & ...  & EA               & 0    & \\
 3750 & HD\,15082       & A5m                   & 1.21988 & 0.017 & ...   & ...  & Det \dag         & 0.08 & {\cite{2010MNRAS.407..507C}} \\
 4290 & HD\,16903       & A3mF2                 & 1.51947 & 0.050 & 0.014 & ...  & EA               & 0    & \\
 4660 & HD\,18597       & A1mF0                 & 2.78071 & 0.681 & 0.522 & ...  & Det              & 0    & XY Cet \\
 5685 & HD\,275604      & A4mF0?                & 7.16050 & 0.555 & 0.149 & 0.51 & EA               & 0    & AB Per \\
 5982 & TYC\,3725-496-1 & A2m?                  & 2.41394 & 0.055 & 0.027 & 0.52 & Ell              & 0.05 & \\
 6720 & HD\,26481       & A2mF2                 & 2.38318 & 0.243 & 0.209 & ...  & Det              & 0    & AE Hor \\
 7310 & HD\,28451       & B9m                   & 6.66372 & 0.328 & 0.154 & ...  & Det              & 0    & ASAS 042815-2306.1 \\
 7730 & HD\,30050       & A5m?                  &39.28272 & 0.817 & 0.073 & 0.65 & Det              & 0    & RZ Eri \\
 8215 & HD\,32239       & A5mF?                 & 8.79590 & 0.234 & 0.099 & ...  & Det              & 0    & \\
 8973 & HD\,243104      & A5m                   & 1.88678 & 0.317 & 0.040 & ...  & EA               & 0.09 & V606 Aur \\
 9237 & HD\,243875      & A2m?                  & 2.85625 & 0.111 & 0.052 & ...  & Det              & 0.35 & \\
 9318 & TYC\,1848-800-1 & A5m?                  &11.11333 & 0.206 & 0.127 & ...  & Det              & 0.01 & \\
 9410 & HD\,36412       & A7mF4                 &16.78729 & 0.552 & 0.124 & 0.49 & Det              & 0    & EY Ori \\
 9458 & HD\,244709      & A3m?                  & 2.25868 & 0.059 & 0.017 & ...  & EA/EB            & 0.14 & \\
10016 & HD\,245819      & A3m                   & 5.43090 & 0.476 & 0.447 & 0.56 & Det              & 0.04 & V1260 Tau \\
10310 & HD\,38303       & A2mA9                 & 1.21672 & 0.295 & 0.050 & ...  & EA               & 0.01 & WZ Pic \\
10326 & HD\,38390       & A3mF4                 & 3.72095 & 0.170 & 0.012 & ...  & EA               & 0    & ASAS 054507-0856.8 \\
10336 & HD\,38453       & A1mF0                 & 2.52600 & 0.203 & 0.025 & ...  & EA               & 0    & ASAS 054602+0212.1 \\
10387 & HD\,247657      & A7m                   & 3.16130 & 0.483 & 0.317 & 0.52 & W UMa            & 0.04 & NSVS 6994211 \\
10689 & HD\,249628      & A2m                   & 1.08374 & 0.140 & 0.097 & 0.59 & W UMa?           & 0.09 & NSVS 7022747 \\
10892 & HD\,250443      & A3m                   & 2.17543 & 0.031 & 0.016 & 0.53 & Ell/Graz         & 0.01 & \\
11100 & HD\,41491       & A1mA5?                & 4.03751 & 0.151 & 0.056 & ...  & Det              & 0    & \\
11387 & HD\,253252      & A4mF1                 & 0.81098 & 0.285 & 0.148 & ...  & W UMa?           & 0.02 & V2787 Ori \\
11470 & HD\,42968       & A0mF1                 & 2.87213 & 0.473 & 0.138 & ...  & EA?/Det?         & 0    & IO CMa \\
14040 & HD\,50992       & A2mA7                 & 1.56695 & 0.105 & 0.045 & ...  & EA/Ell           & 0.24 & \\
14850 & HD\,54011A      & A1mF0?                & 3.97948 & 0.084 & 0.008 & ...  & Det \dag         & 0.02 & \\
15034 & HD\,55228       & F2m Sr                & 7.53921 & 0.172 & 0.095 & ...  & Det?             & 0    & V422 Gem \\
15190 & HD\,55822A      & A3mF5                 & 5.12290 & 0.062 & ...   & ...  & Det \dag         & 0    & \\
15445 & HD\,56587       & A3mF2                 & 5.76059 & 0.520 & 0.462 & ...  & Det              & 0    & V339 Gem \\
18505 & HD\,67093       & A3mF0                 & 4.33586 & 0.340 & 0.319 & ...  & Det              & 0    & V871 Mon \\
21400 & HD\,76320       & A2m                   & 7.77292 & 0.193 & 0.009 & ...  & Det \dag         & 0    & \\
22860 & HD\,80343       & A3mA9                 & 7.90058 & 0.076 & 0.066 & 0.66 & Det              & 0    & \\
25020 & HD\,87374       & A0m?                  & 6.62845 & 0.009 & ...   &...   & Det \dag         & 0.05 & \\
25070 & HD\,87450       & A1mF2                 & 6.71489 & 0.228 & 0.214 & 0.58 & Det              & 0    & ASAS 100421-3319.0 \\
25880 & HD\,90029       & A5m $\delta$\,Del      & 9.86030 & 0.074 & 0.052 & ...  & Det              & 0    & BY Ant \\
28850 & HD\,100376      & F0m? $\delta$\,Del?    & 1.64361 & 0.046 & 0.046 & ...  & Ell/Cont/grazing & 0    & ASAS 113257-2737.4 \\
29290 & HD\,101681      & A3m?                  & 3.29220 & 0.188 & 0.179 & ...  & Det \dag         & 0.05 & ASAS 114149-4229.5 \\
30090 & HD\,104120      & A3mF2                 & 4.34862 & 0.155 & 0.155 & ...  & Det \dag EA?     & 0.01 & \\
30110 & HD\,104186      & A5m?                  & 4.31449 & 0.029 & ...   & ...  & Det              & 0    & \\
30457 & HD\,105376      & A2mA8                 &11.94200 & 0.069 & ...   & ...  & Det \dag         & 0    & \\
30650 & HD\,106046      & A2mF0                 &18.12101 & 0.168 & ...   & ...  & Det \dag         & 0    & \\
30820 & HD\,106546      & A0m                   & 2.87025 & 0.025 & 0.006 & ...  & EA               & 0    & \\
34770 & HD\,120777      & A2mF0                 & 2.54163 & 0.020 & 0.005 & 0.57 & EA  & 0.03 & \\
35000 & HD\,121788      & A2 Sr Cr or Am?       &10.28606 & 0.146 & ...   & ...  & Det \dag         & 0    & ASAS 135817-3004.5 \\
36660 & HD\,128806      & A1mF2                 &16.36534 & 0.559 & 0.257 & 0.44 & Det              & 0.02 & ASAS 143944-2837.2 \\
37220 & HD\,130922      & F5m?                  & 5.79311 & 0.120 & 0.117 & ...  & Det \dag         & 0.34 & \\
37610 & HD\,132515A     & F8 Sr or $\delta$\,Del & 3.23869 & 0.236 & 0.314 & ...  & Det              & 0.18 & IU Lup \\                                             
38180 & HD\,134477      & A1mA6                 & 6.14445 & 0.075 & 0.035 & ...  & EA               & 0    & OY Lup \\
38500 & HD\,135492      & A2mA9                 & 3.99382 & 0.090 & 0.020 & ...  & Det              & 0.04 & \\
40350 & HD\,142232      & A3mF2                 & 7.06875 & 0.103 & ...   & ...  & Det \dag         & 0    & \\
40780 & HD\,143926      & A5mF0                 & 6.93480 & 0.175 & 0.170 & ...  & Det              & 0    & \\
40910 & HD\,144396      & A1mF0                 &11.11629 & 0.353 & 0.349 & 0.47 & Det              & 0.11 & V1046 Sco \\
42906 & HD\,151604      & A0m                   &19.69874 & 0.285 & ...   & ...  & Det \dag         & 0.04 & V916 Her \\
44140 & HD\,156965      & A5mA9                 & 2.05984 & 0.630 & 0.403 & ...  & Det              & 0    & TX Her \\
49380 & HD\,177022      & F4m?                  & 5.02043 & 0.096 & 0.049 & 0.56 & Det              & 0.27 & \\
51506 & HD\,186753      & A2mF0?                & 1.91955 & 0.018 & 0.004 & 0.44 & Det         & 0    & {\cite{2009A&A...508..391B}} \\
56310 & HD\,201964      & A2m                   & 2.69592 & 0.413 & 0.354 & ...  & Det              & 0.03 & DG Mic \\
56830 & HD\,204038      & A3mF0                 & 0.78582 & 0.321 & 0.286 & ...  & Ell/Cont/grazing & 0.03 & V1073 Cyg \\
57845 & HD\,208090      & A2m $\delta$\,Del?     & 2.44660 & 0.176 & 0.048 & ...  & EA?              & 0    & {\cite{2011MNRAS.416.2477W}} \\
58170 & HD\,209147      & A2mA4                 & 1.60471 & 0.947 & 0.338 & ...  & Det              & 0    & CM Lac \\
58256 & HD\,209385      & A3mF3                 & 2.96733 & 0.140 & 0.062 & ...  & Det              & 0.05 & \\
59780 & HD\,216429      & A1mA8?                & 7.35140 & 0.592 & 0.498 & 0.51 & Det              & 0.04 & V364 Lac \\
60640 & HD\,221184      & A5m?                  & 5.46091 & 1.299 & 0.081 & ...  & EA               & 0    & AN Tuc \\
61280 & TYC\,6408-989-1 & A4m or A5 Sr?         & 0.47080 & 0.373 & 0.178 & ...  & Ell/Cont/grazing & 0    & ASAS J235103-1904.5 \\
\hline
\end{tabular}
\end{table*}

\begin{table}[h]
\caption{Suspected eclipsing Am binaries.}
\label{table:Am Binaries No Period}
\scriptsize
\begin{tabular}{llll}
\hline\hline
Renson  & Name      & Sp. Type          & Notes\\\hline
7360    & HD\,28617  & A0mA5 ?           & Eclipse, JD 4396.64, depth 0.17\,{mag.} \\
9701    & HD\,245224 & A2m               & Egress, JD 4083.50, depth 0.1\,{mag.} \\
36950   & HD\,129575 & F0m $\delta$\,Del & Possible egress, JD 3891.20, depth 0.2\,{mag.} \\
61470   & HD\,224401 & A4mF2             & Egress, JD 5399.40, depth 0.1\,{mag.} \\
\hline
\end{tabular}
\tablefoot{Dates are given as JD-2450000}
\end{table}

\subsection{Cross-checking with AAVSO}

In order to check whether any known eclipsing systems had been missed, a
cross-check with AAVSO \citep{2006SASS...25...47W} was performed. All but two of
the known systems were recovered in the WASP data. The first system,
Renson\,5740 (BD+44 765) is listed as an XO false positive
\citep{2010ApJS..189..134P} with an eclipse depth of 0.018\,{mag.} and duration of
3.37\,hours. There is only a single, rather noisy, WASP light curve and folding on
their ephemeris, shows no sign of any transits. Furthermore, only two transits
would have occurred within the WASP light curve, which is less than the minimum 3
required for detection of a period. Thus, even if the eclipses had been found,
the period would have been unknown. The other system, Renson\,34764 (HD\,120727)
is a suspected eclipsing system \citep{1990ApJS...74..225H}. However, the 3 good
quality WASP light curves do not show any evidence of eclipses. Hence, we
conclude that this is not an eclipsing system.

\subsection{Systems also showing pulsations}

In \cite{2011A&A...535A...3S} we found that approximately 14\% of Am stars
pulsate with amplitudes $\ga1$\,{mmag.}. Hence, we might expect some of the binary
systems to show evidence of pulsations. A search for pulsations was undertaken
using the residuals to the fitted light curves.

Two of the binary systems already have known pulsations. Renson\,3750
(HD\,15082) was found to exhibit $\delta$ Scuti pulsations at the
milli-magnitude level \citep{2011A&A...526L..10H}, but these are not detectable
in the WASP data. Renson\,5685 (HD\,275604; AB Per) was reported to have
10\,{mmag.} pulsations in the $B$ band with a frequency of 5.106\,d$^{-1}$
\citep{2003A&A...405..231K}. There is a $\sim$6\,{mmag.} peak in the periodogram
of the residuals for the multi-season combined light curve with a frequency of
around 5.116\,d$^{-1}$, which confirms the previous detection.

Of the remaining systems, Renson\,8973 (HD\,243104; V606 Aur) was found to
clearly exhibit pulsations. This system has 11.9\,{mmag.} $\delta$ Scuti-type
pulsations with a frequency of 23.572\,d$^{-1}$. Another system, Renson\,10310
(HD\, 38303; WZ Pic) shows 1.5\,{mmag.} pulsations with a frequency of
22.783\,d$^{-1}$, but individual seasons show this period at $\pm$1\,d$^{-1}$
aliases. Another system, Renson\,30110 (HD\,104186) shows some evidence for
excess power in the individual light curves at the 1$\sim$2\,{mmag.} level
around 10\,d$^{-1}$. However, none of them yield consistent frequencies. Thus we
conclude that the hints of pulsations in Renson\,30110 are probably spurious.

\section{Spectroscopic observations}

Spectroscopic observations of five of the Am binary systems were obtained at the
2.15m telescope at the Complejo Astron\'omico el Leoncito (CASLEO) on the nights
between the 2009 June 12 and the 2009 June 18. A Tektronik 1024$\times$1024 CCD
and the REOSC echelle spectrograph with the either the grating 580
(400\,l\,mm$^{-1}$) or grating 260 (600\,l\,mm$^{-1}$) as detailed in
Table\,\ref{obs-log}. The spectral resolution was 25600 and the integration
times were 1800s.

\begin{table}
\caption{Spectrograph setting for each night}
\label{obs-log}
\begin{tabular}{llll} \hline\hline
Day & Grating & Angle & Wavelength Range \\ \hline
12 & 260 & 8\degr 25\arcmin  & 3663--5264 \\
13 & 260 & 9\degr 50\arcmin  & 4396--5968 \\
14 & 260 & 9\degr 10\arcmin  & 4053--4943 \\
15 & 580 & 7\degr 00\arcmin  & 3940--6305 \\
16 & 580 & 7\degr 00\arcmin  & 3940--6305 \\
17 & 260 & 10\degr 30\arcmin & 4888--6485 \\
18 & 260 & 9\degr 10\arcmin  & 4053--4943 \\ \hline
\end{tabular}
\end{table}

Data reduction was performed using \textsc{iraf}
\citep{1986SPIE..627..733T,1993ASPC...52..173T}. Master bias and flat field
frames were obtained by combining sets of 50 individual images. The stellar
spectra were bias subtracted and divided by the normalized master flat field.
They were then cleaned for cosmic rays and scattered light corrected. The
echelle orders were extracted to produce spectra for each individual order and
wavelength calibrated using ThAr lamp spectra.

Radial velocities were obtained by cross-correlation with synthetic spectra
generated using \textsc{uclsyn} \citep{1988eaa..conf...32S}. The heliocentric
values are given in Table\,\ref{RV}.

\begin{table}
\caption{Heliocentric radial velocity measurements for five Am binary systems.
The uncertainty in RV is 5\,km\,s$^{-1}$.}
\label{RV}
\begin{tabular}{llll}\\ \hline\hline
HJD-2450000 & RV$_1$ & RV$_2$ & RV$_3$ \\ \hline
\multicolumn{4}{l}{Renson\,25070 (HD\,87450)}\\
4995.482959 &   +4 & ...  & \\
4996.475654 &  +88 &$-$80 & \\
4998.480418 &   +6 & ...  & \\
4999.490138 &$-$60 &  +64 & \\
5000.501385 &$-$68 &  +75 & \\
5001.493102 &$-$36 &  +44 & \\ \hline
\multicolumn{4}{l}{Renson\,34770 (HD\,120777)}\\
4995.549946 &$-$37 &       & \\
4996.565241 &$-$15 &       & \\
4998.534940 &$-$18 &       & \\
5000.559216 &$-$30 &       & \\
5001.576971 &$-$12 &       & \\ \hline
\multicolumn{4}{l}{Renson\,36660 (HD\,128806)}\\
4995.605844 &$-$57 &  +20 & \\
4996.619294 &$-$36 &$-$1  & \\
4997.589355 &$-$19 & ...  & \\
4998.650717 &$-$21 & ...  & \\
4999.570940 &$-$16 & ...  & \\
5001.673912 &$-$1  &$-$51 & \\ \hline
\multicolumn{4}{l}{Renson\,49380 (HD\,177022)}\\
4995.7364532 & +138  &$-$198 &$-$45 \\
4996.7481885 &  +35  &$-$142 &$-$44 \\
4998.7799882 &$-$95  &   +22 &$-$48 \\
4999.7443681 &$-$103 &   +24 &$-$46 \\
5001.7471764 &   +35 &$-$135 &$-$41 \\ \hline
\multicolumn{4}{l}{Renson\,51506 (HD\,186753)}\\
4995.810117 &$-$2  & & \\
4997.798468 &$-$7  & & \\
4998.851621 &$-$51 & & \\
4999.819895 &$-$17 & & \\
5001.855602 &$-$16 & & \\
\hline
\end{tabular}
\tablefoot{\\HD\,177022 is a visual double comprising two 10.7 stars separated by
0.2{\arcsec}, RV$_3$ gives measurements for the `stationary' component.}
\end{table}

\section{Spectroscopic orbits}                                                                                                            \label{sec:model}
\label{Data_modelling}

The light curves of many of the systems are extensively covered by SuperWASP
observations, making a preliminary analysis of individual objects worthwhile. We
also possess radial velocity (RV) measurements for 5 systems, opening the
possibility of obtaining a full set of physical properties.

For the light curve analysis we chose to use the \textsc{jktebop} code
\citep{2004MNRAS.351.1277S,2008MNRAS.386.1644S}, which is suitable for detached eclipsing binaries
(dEBs) with only moderately distorted stars. \textsc{jktebop} has recently been
extended to include the simultaneous fitting of one light curve and RVs for both
components \citep{2013A&A...557A.119S}. The sizes of the primary and secondary star are
parametrised using the fractional radii, $r_{\rm A} = \frac{R_{\rm A}}{a}$ and
$r_{\rm B} = \frac{R_{\rm B}}{a}$, where $a$ is the orbital semi-major axis and
$R_{\rm A}$ and $R_{\rm B}$ are the true radii of the stars. The main parameters
of the fit are the sum and ratio of the fractional radii, $r_{\rm A} + r_{\rm
B}$ and $k = \frac{r_{\rm B}}{r_{\rm A}} = \frac{R_{\rm B}}{R_{\rm A}}$, the
orbital inclination $i$, the central surface brightness ratio of the two stars
$J$, the orbital period {$P_{\rm orb}$} and the time of primary mid-eclipse $T_0$.

In the cases of eccentric orbits the orbital eccentricity, $e$, and longitude of
periastron, $\omega$, were included using the combination terms $e\cos\omega$
and $e\sin\omega$. The value of $e\cos\omega$ is closely related to the orbital
phase at which the secondary eclipse occurs, so is usually measured precisely.
On the other hand, $e\sin\omega$ is less well tied down as it primarily
determines the ratio of the durations of the eclipses. The precision of the
measurements of $e\cos\omega$ and $e\sin\omega$ are significantly improved when
RVs can be included as well as light curves in a solution. When RVs were
available we also fitted for one or both of the velocity amplitudes, $K_{\rm A}$
and $K_{\rm B}$, as well as the systemic velocities of the star.

We checked for contaminating `third light', $L_3$, from additional stars in the
same point spread function as our target stars. The value of $L_3$ was set to
zero unless there was clear evidence of its existence. We also fitted for the
out-of-transit magnitudes of the stars and in some cases the size of the
reflection effect. Limb darkening was implemented using the linear law with
appropriate coefficients, and reasonable choices of the coefficients have a
negligible effect on our results. 

As a first step for each object, we determined an initial orbital ephemeris
manually and then ran preliminary fits to its light curve alone.
An iterative 3$\sigma$ clip was used to remove discrepant data points affected
by weather or instrumental problems. We then assigned the
same measurement error to every data point of such a size as to yield a reduced
$\chi^2$ value of unity for the best fit. The radial velocities were then added
into the solution and their error bars were adjusted to yield reduced $\chi^2$
values near unity for individual data sets. 

Uncertainties in the deduced parameters were assessed using Monte Carlo and
residual-permutation simulations
\citep{2004MNRAS.351.1277S,2008MNRAS.386.1644S}. 1$\sigma$ error bars were
estimated by marginalising over the parameter distributions for these
simulations. In line with previous experience with SuperWASP data we find that
the residual-permutation uncertainties are typically twice as large as the Monte
Carlo uncertainties, and we quote the larger of the two alternatives for each
measured parameter.

\begin{table*}
\label{absdim}
\scriptsize
\caption{Measured properties of the systems with RV measurements.}
\centering
\begin{tabular}{llllll} \hline \hline
Parameter               & Renson\,25070           & Renson\,34770           & Renson\,36660              & Renson\,49380           & Renson\,51506           \\ \hline
{$P_{\rm orb}$} (d)     & 6.714890 $\pm$ 0.000011 & 2.541648 $\pm$ 0.000016 & 16.36534 $\pm$ 0.00011     & 5.02068 $\pm$ 0.00003   & 1.919549 $\pm$ 0.000019 \\
$T_0$ (HJD-2450000)     & 4145.5389 $\pm$ 0.0007  & 4546.9679 $\pm$ 0.0021  & 4614.3659 $\pm$ 0.0015     & 3903.5220 $\pm$ 0.0015  & 4272.4863 $\pm$ 0.0034  \\
$r_{\rm A} + r_{\rm B}$ & 0.2006 $\pm$ 0.0005     & 0.255 $\pm$ 0.005       & 0.0986 $\pm$ 0.0016        & 0.174 $\pm$ 0.014       & 0.390 $\pm$ 0.038       \\
$k$                     & {1.0 fixed}             & 0.0987 $\pm$ 0.0022     & 1.038 $\pm$ 0.05           & 0.9 fixed               & 0.11157 $\pm$ 0.0084     \\
$i$                     & 83.90 $\pm$ 0.03        & 90.0 $\pm$ 1.5          & 88.13 $\pm$ 0.07           & 84.7 $\pm$ 1.6          & 75.0 $\pm$ 3.8          \\
$J$                     & 1.05 $\pm$ 0.02         & 0.070 $\pm$ 0.022       & 0.90 $\pm$ 0.07            & 0.78 $\pm$ 0.32         & 0.187 $\pm$ 0.004       \\
$e\cos\omega$           & 0.131 $\pm$ 0.001       & 0.106 $\pm$ 0.003       & $-$0.082 $\pm$ 0.001       & 0.085 $\pm$ 0.001       & $-$0.092 $\pm$ 0.024    \\
$e\sin\omega$           & 0.036 $\pm$ 0.006       & 0.224 fixed             & 0.420 $\pm$ 0.014          & 0.117 $\pm$ 0.032       & $-$0.110 $\pm$ 0.130      \\
$K_{\rm A}$ (km\,s$^{-1}$) & 87 $\pm$ 4              & 15 $\pm$ 2              & 49 $\pm$ 9                 & 76.5 $\pm$ 2.2          & 23.2 $\pm$ 2.5          \\
$K_{\rm B}$ (km\,s$^{-1}$) & 86 $\pm$ 5              &                         & 65 $\pm$ 7                 & 92.2 $\pm$ 2.5          &                         \\ \hline
Light ratio             & 1.05 $\pm$ 0.02         & 0.00068 $\pm$ 0.00003   & 0.97 $\pm$ 0.10            & 0.63 $\pm$ 0.26         & 0.0055 $\pm$ 0.0034     \\
$e$                     & 0.1358 $\pm$ 0.0016     & 0.2476 $\pm$ 0.0013     & 0.428 $\pm$ 0.013          & 0.145 $\pm$ 0.026       & 0.15 $\pm$ 0.10         \\
$\omega$ (degrees)      & 16 $\pm$ 3              & 65 $\pm$ 2              & 101.1 $\pm$ 0.4            & 54 $\pm$ 8              & 129 $\pm$ 27            \\ \hline
$a$ ($R_\sun$)          & 23.0 $\pm$ 1.1          &                         & 33$^1$ $\pm$ 5             & 16.6 $\pm$ 0.5          &                         \\
$M_{\rm A}$ ($M_\sun$)  & 1.8 $\pm$ 0.3           &                         & 1.0$^1$ $\pm$ 0.4          & 1.34 $\pm$ 0.11         & 1.8$^3$                 \\
$M_{\rm B}$ ($M_\sun$)  & 1.8 $\pm$ 0.2           &                         & 0.8$^1$ $\pm$ 0.4          & 1.11 $\pm$ 0.09         & 0.22$^3$                \\
$R_{\rm A}$ ($R_\sun$)  & 2.31 $\pm$ 0.10         &                         & 1.61$^1$ $\pm$ 0.23        & 1.52 $\pm$ 0.13         & 2.9$^3$                 \\
$R_{\rm B}$ ($R_\sun$)  & 2.31 $\pm$ 0.10         &                         & 1.67$^1$ $\pm$ 0.22        & 1.37 $\pm$ 0.12         & 0.33$^3$                \\
$\log g_{\rm A}$ (cgs)  & 3.97 $\pm$ 0.02         &                         & 4.05$^2$ $\pm$ 0.05        & 4.20 $\pm$ 0.06         &                         \\
$\log g_{\rm B}$ (cgs)  & 3.97 $\pm$ 0.02         &                         & 3.90$^2$ $\pm$ 0.08        & 4.21 $\pm$ 0.06         &                         \\ \hline 
\end{tabular} 
\tablefoot{$^1$ These numbers are likely to be too low due to spectral line blending. 
\newline $^2$ These numbers are likely to be too high due to spectral line blending.
\newline $^3$ Inferred using theoretical stellar models to obtain the mass of the primary star.}
\end{table*}

\subsection{Renson\,25070 (HD\,87450)}

\begin{figure}
\includegraphics[width=\columnwidth,angle=0]{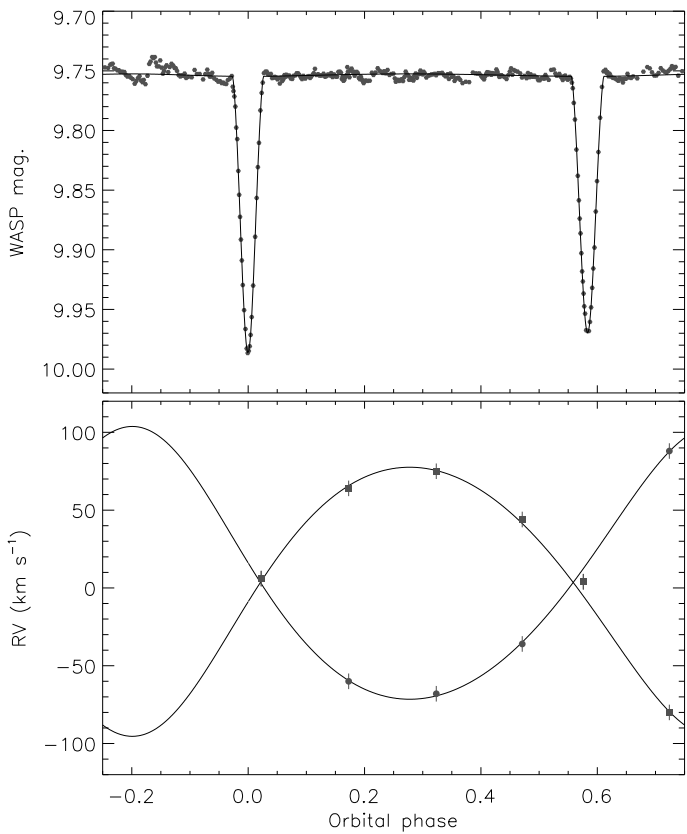}
\caption{Observed light and RV curves of Renson\,25070 (points) 
compared to the best fit found using \textsc{jktebop} (lines).
For presentation purposes only, the WASP light curve
has been binned into 400 phase bins.}
\label{fig:25070}
\end{figure}

This object shows eclipses 0.25\,mag deep on an orbital period of 6.7\,d. The
secondary eclipse is almost as deep as the primary, showing that the two stars
have almost the same surface brightness and are probably very similar
stars. The stars are well-detached and in a mildly eccentric orbit: secondary
eclipse occurs at phase 0.583. We obtained six spectra of Renson\,25070 on
almost-successive nights. Two were taken when the velocities of the stars
were similar and their spectral lines were not resolved, but the remaining four
were taken when the lines were nicely separated. All six RVs were used for each
star, with the ones near conjunction down-weighted by a factor of ten. A total
of 18\,137 data points are included in the light curve.

The partial eclipses combined with two similar stars led to a solution which was
poorly defined, so we fixed the ratio of the radii to be $k = 1$ for our final
solution. The measured mass ratio is consistent with unity, which supports this
decision. Both stars have a mass of 1.8\,{$M_\sun$} and a radius of
2.3\,{$R_\sun$}, so are slightly evolved. The fits to the light and RV curves
are shown in Fig.\,\ref{fig:25070} and the fitted parameters are given in
Table\,\ref{absdim}. The masses, radii and surface gravities have the symbols
$M_{\rm A}$ and $M_{\rm B}$, $R_{\rm A}$ and $R_{\rm B}$, and $\log g_{\rm A}$
and $\log g_{\rm B}$, respectively. Note that the uncertainties are
underestimated because we have imposed the constraint $k=1$ on the solution.

\subsection{Renson\,34770 (HD\,120777)}

\begin{figure}
\includegraphics[width=\columnwidth,angle=0]{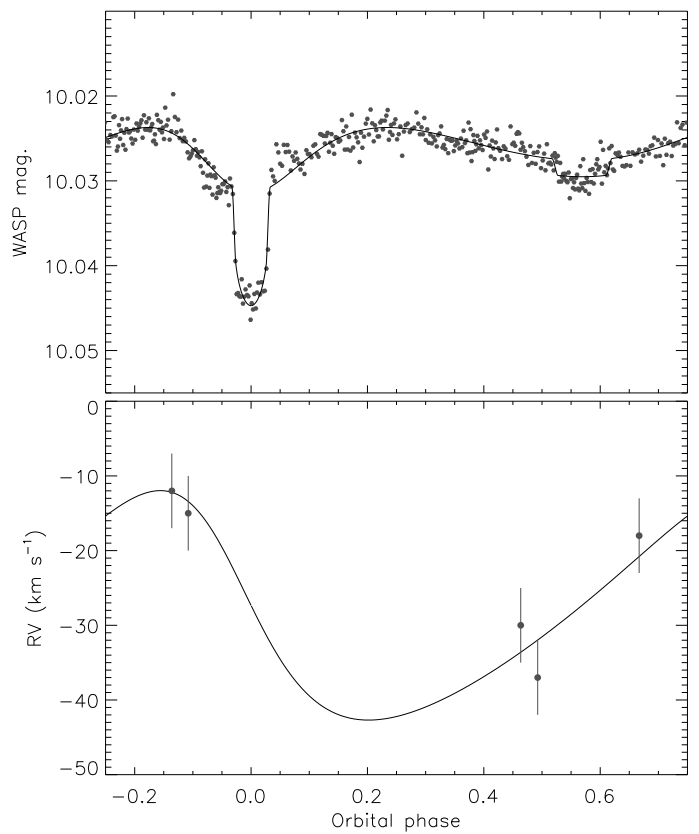}
\caption{Observed light and RV curves of Renson\,34770 (points) 
compared to the best fit found using \textsc{jktebop} (lines).
For presentation purposes only, the WASP light curve
has been binned into 400 phase bins.}
\label{fig:34770}
\end{figure}

Renson\,34770 shows shallow eclipses on a period of 2.5\,d. The primary eclipse
is securely detected with a depth of 0.014\,{mag.}, but the secondary eclipse is
only speculatively detected with a depth of 0.002\,{mag.}. The orbit is moderately
eccentric and secondary eclipse occurs at phase 0.569. The secondary star is a
low-mass object with a radius ten times smaller than that of the primary. Five
RVs were measured for the primary star, but the secondary could not be detected
in the spectrum. A joint fit to the light curve and RVs of the primary star was
poorly determined, so we fixed $e\cos\omega = 0.224$ to obtain a reasonable
solution indicative of the properties of the system. This solution is shown in
Fig.\,\ref{fig:34770} and the fitted parameters are in
Table\,\ref{absdim}.

A definitive analysis will require high-quality photometry to measure the depth
and shape of the primary and specifically the secondary eclipse. Whilst the
SuperWASP light curve contains 13\,144 data points, they have an rms of 7\,mmag
versus the fitted model. The secondary component is not of planetary mass -- it
is massive enough to induce tidal deformation of the primary star which
manifests as ellipsoidal variations easily detectable in the SuperWASP light
curve.

\subsection{Renson\,36660 (HD\,128806)}

\begin{figure}
\includegraphics[width=\columnwidth,angle=0]{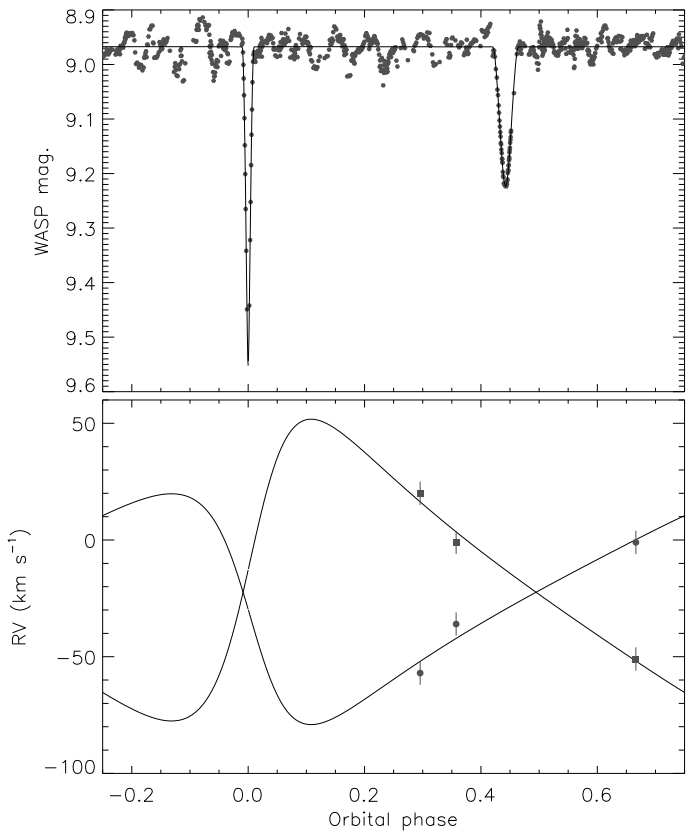}
\caption{Observed light and RV curves of Renson\,36660 (points) 
compared to the best fit found using \textsc{jktebop} (lines).
For presentation purposes only, the WASP light curve
has been binned into 1000 phase bins.}
\label{fig:36660}
\end{figure}

The SuperWASP light curve of Renson\,36660 has 9089 data points and shows
significant systematic trends due to the brightness of the system putting it
near the saturation limit. Its orbit is eccentric -- the secondary eclipse is
much longer than the primary and occurs at phase 0.442 -- with a period of
16.4\,d. The eclipses are partial and are deep at 0.6\,mag and 0.3\,mag,
respectively. We obtained six spectra and were able to measure RVs from
three of them. These RVs were included in
the fit (Fig.\,\ref{fig:36660}), yielding the full physical properties of the
system (Table\,\ref{absdim}).

The masses we find (1.0 and 0.8\,$M_\sun$) are significantly too low for
the spectral type of the system (A1m), an effect which is likely due to
spectral line blending \citep[e.g.][]{1975A&A....44..355A}. Three of our spectra
show fully blended lines and were not measured for RV. Two more of the spectra
suffer from significant line blending, and only one spectrum (that at phase 0.3)
has cross-correlation function peaks from the two stars which are clearly
separated. An investigation and mitigation of this problem could be achieved by
a technique such as spectral disentangling \citep{1994A&A...281..286S}, but this
requires significantly more spectra than currently available so is beyond the
scope of the present work.

We crudely simulated the effects of line blending by moving each of the
four RVs from the blended spectra by 5\,km\,s$^{-1}$ {\it away} from the
systemic velocity. The resulting solution gave lower RV residuals and masses of
1.5 and 1.2\,$M_\sun$, showing that a modest amount of blending can easily move
the measured masses to more reasonable values. For the current work we present
our solution with the measured RVs rather than those with an arbitrary
correction for blending, and caution that much more extensive observational
material is required to obtain the properties of the system reliably.

The physical properties of the stars in our preliminary solution were
very uncertain, in particular the light ratio between the two objects. The main
problem was the well-known degeneracy between $k$ and $J$ measured from the deep
but partial eclipses, exacerbated by correlations with $e\sin\omega$ for this
eccentric system \citep[e.g.][]{1981AJ.....86..102P}. We therefore measured a
spectroscopic light ratio of $0.8 \pm 0.2$ from the line strengths in the
spectrum which shows well-separated lines, and applied it to the
\textsc{jktebop} solution using the method of \citet{2007A&A...467.1215S}. This
makes the radii of the two stars much more precise, but they will still be
too small because the line blending causes an underestimation of the orbital
semimajor axis as well as the stellar masses. The correlated noise in the
SuperWASP light curve results in large uncertainties in the measured photometric
parameters.

\subsection{Renson\,49380 (HD\,177022)}

\begin{figure}
\includegraphics[width=\columnwidth,angle=0]{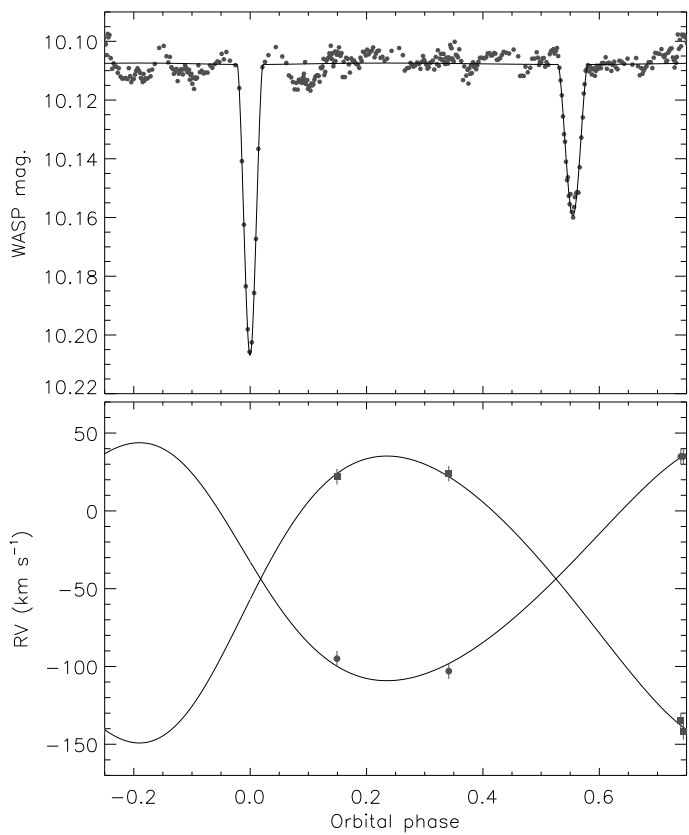}
\caption{Observed light and RV curves of Renson\,49380 (points) 
compared to the best fit found using \textsc{jktebop} (lines).
For presentation purposes only, the WASP light curve
has been binned into 400 phase bins.}
\label{fig:49380}
\end{figure}

The SuperWASP light curve of this object (8029 data points) shows shallow
eclipses of depth 0.10 and 0.05\,{mag.}, respectively. The 5.0\,d orbit is eccentric, and
secondary minimum occurs at phase 0.555.  It is in a crowded field and many
fainter stars are positioned inside the photometric aperture. We therefore
allowed for third light when fitting the light curve, finding a value of $L_3 =
0.64 \pm 0.06$. 

We obtained four spectra of Renson\,49380, all taken when the velocity separation
of the two stars was at least 100\,km\,s$^{-1}$. The best solution to all data has an rms
residual of 8\,{mmag.} for the photometry and 3\,km\,s$^{-1}$ for the RVs. This is plotted
in Fig.\,\ref{fig:49380} and the parameters are given in
Table\,\ref{absdim}. The parameters of a free fit are poorly defined
because $L_3$ is strongly correlated with $k$, so we set $k$ to a reasonable
value of 0.9 to obtain a nominal solution.

The stars have masses of 1.3 and 1.1\,{$M_\sun$} and radii of 1.5 and
1.4\,{$R_\sun$}. These numbers are rather modest, but agree with the
$T_\mathrm{eff}$ = 6000\,K suggested by colour indices of the system. The stars are slightly too
late-type to be Am stars, so this could be a case of misclassification of a
composite spectrum as a metal-rich spectrum.

\subsection{Renson\,51506 (HD\,186753)}

\begin{figure}
\includegraphics[width=\columnwidth,angle=0]{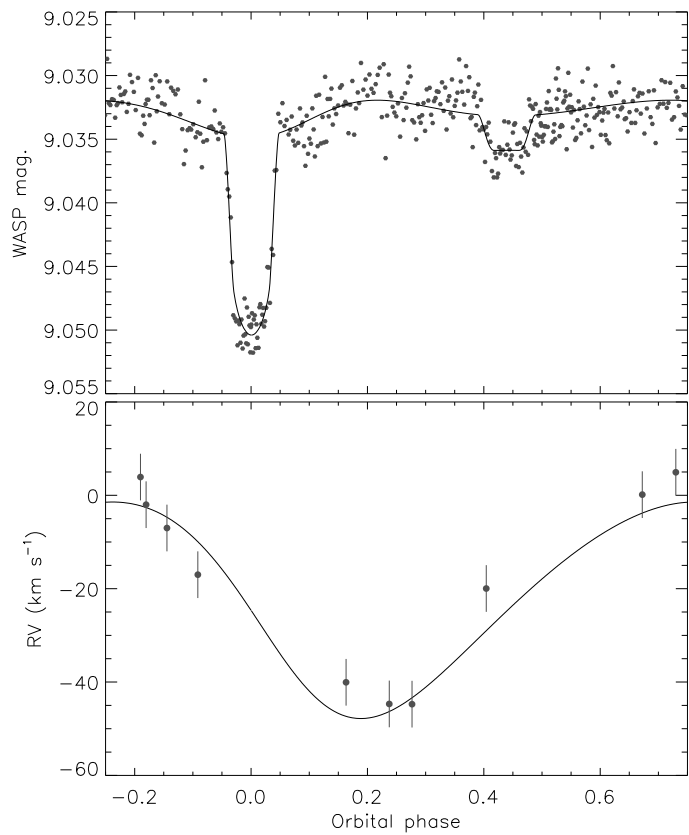}
\caption{Observed light and RV curves of Renson\,51506 (points) 
compared to the best fit found using \textsc{jktebop} (lines).
For presentation purposes only, the WASP light curve
has been binned into 400 phase bins.}
\label{fig:51506}
\end{figure}

This star was identified as a dEB consisting of an A\,star and an M\,star by
\citet{2009A&A...508..391B}, who presented eight RV measurements of the A\,star and
combined these with the SuperWASP light curve to obtain the physical properties
of the system. We have revisited this system because further SuperWASP data (now
totalling 9896 points) and five more spectra are available. The light curve
shows a clear detection of the secondary minimum (Fig.\,\ref{fig:51506}), with a
significance of 3.4$\sigma$, at orbital phase 0.441. 

The RVs of Renson\,51506 are relatively poorly defined, due to the high
rotational velocity of the primary star ($v \sin i$ = 65 $\pm$ 5\,km\,s$^{-1}$;
\citealt{2009A&A...508..391B}). We rejected the single HARPS measurement from
\citet{2009A&A...508..391B}, and also two of our own measurements which
were discrepant with both the best fit and a CORALIE RV obtained at the
same orbital phase. This allowed us to obtain a determinate solution of the
light and RV curves (Fig.\,\ref{fig:51506} and Table\,\ref{absdim}). Those
parameters also measured by \citet{2009A&A...508..391B} are all within 1$\sigma$
of the values we find. 

Whilst we lack RVs of the secondary star, we were able to estimate the
full physical properties of the system by finding the $K_{\rm B}$ which
reproduces the primary star mass of $M_{\rm A} = 1.794$\,{$M_\sun$} obtained by
\citet{2009A&A...508..391B} from interpolation in theoretical models. Adopting
$K_{\rm B} = 187.9$\,km\,s$^{-1}$ gives masses of 1.8 and 0.22\,{$M_\sun$} and
radii of 2.9 and 0.33\,{$R_\sun$} for the two stars. This is unsurprisingly in
good agreement with the values found by \citet{2009A&A...508..391B}. The
secondary star is of very low mass and has a radius too large for theoretical
predictions; such discrepancies have been recorded many times in the past
\citep[e.g.][]{1973A&A....26..437H,2006Ap&SS.304...89R,2007ApJ...660..732L}.
Near-infrared spectroscopy of the Renson\,51506 system could allow measurement
of the orbital motion of the secondary star, which together with the existing
light curve would yield the full physical properties of the system without
reliance on theoretical models. The secondary star could then be used as a probe
of the radius discrepancy in the crucial 0.2\,{$M_\sun$} mass regime.

\subsection{Other objects}

RVs are not available for the other Am-type EBs studied in this work. We modelled
the light curves of these objects with the primary aim of determining reliable
orbital periods to facilitate population studies and follow-up observations.
After obtaining preliminary solutions we performed iterative 3$\sigma$ rejection
of discrepant points to arrive at a light curve fit for more detailed analysis. 

A small fraction of the systems show strong tidal interactions which deform the
stars beyond the limits of applicability of the \textsc{jktebop} code. Reliable
solutions could be obtained by the use of a more sophisticated model, such as
implemented in the Wilson-Devinney code \citep{1971ApJ...166..605W}, at the
expense of much greater effort and calculation time. This work is beyond the
scope of the paper; in these cases \textsc{jktebop} is still capable of returning
the reliable orbital ephemerides which are our primary goal when modelling the
light curves.

\section{Detection probability}
\label{DetProb}

In order to assess whether the observed fraction of eclipsing Am stars is
consistent with the expected fraction of Am binaries, we need to determine the
detection probability. Of the 1742 stars in our sample, 282 have
$uvby\beta$ photometry which gives an average $T_\mathrm{eff}$ = 7520$\pm$580\,K
using the calibration of \cite{1985MNRAS.217..305M}. Hence, in the following,
we assume that a typical Am star is around $T_\mathrm{eff}$ = 7500\,K, with
$R=1.7\,R_{\sun}$, $M=1.7\,M_{\sun}$ and $L = 7\,L_{\sun}$. We will consider two
scenarios; two identical $1.7\,R_{\sun}$ stars and the case of a dark companion
with radius $\sim$0.2\,$R_{\sun}$. In both cases, we assume that the orbits are
circular.

\subsection{Eclipsing probability}
\label{ecl_prob}

The probability of an eclipse being seen
from the Earth is given by
\[ p_{\rm eclipse} = \frac{R_1 + R_2}{a}, \]
where $R_1$ and $R_2$ are the stellar radii of the two
stars and $a$ is orbital separation in same units.
The above criterion is purely geometric and does not take into
account the effects of limb-darkening or noise on the detection of shallow
eclipses. Hence, the maximum sky-projected separation of the centres of the two
stars which will lead to a detectable eclipse will be less than $R_1 + R_2$ by
an amount $\delta R$. The probability of detectable eclipse is therefore,
\[ p_{\rm eclipse} = \frac{R_1 + R_2 - \delta R}{a}. \]
Assuming a linear limb-darkening law with $\epsilon = 0.6$ and a
minimum detectable eclipse depth of 0.01\,{mag.}, we get
$\delta R$=0.28\,$R_1$ and $\delta R$=0.15\,$R_1$, for a
0.1\,$R_1$ dark companion and two identical stars, respectively. Using
Kepler's 3rd Law, we get
\[ p_{\rm eclipse} =  0.23756 \frac{(R_1 + R_2 - \delta R)}{\sqrt[3]{P_{\rm orb}^2 (M_1+M_2)}},\]
where $R_1$, $R_2$ and $\delta R$ are in
$R_{\sun}$, $M_1$ and $M_2$ are in $M_{\sun}$ and $P_{\rm orb}$ in days.
An uncertainty of $\pm$0.3$R_{\sun}$ and $\pm$0.3$M_{\sun}$ in stellar
radii and masses, yields an uncertainty in $p_{\rm eclipse}$ of approximately
10\%, with the uncertainty being dominated by that on the stellar radius.

In the above we have assumed that the orbits are circular. However, a 
significant fraction of short-period Am binaries have eccentric orbits
\citep{2004ASPC..318..297N}, but most of the systems with periods less than 10
days have $e < 0.3$. Eccentricity has the effect of increasing the probability
that at least one eclipse per orbital period might occur by a factor of
$(1-e^2)^{-1}$. For example, the eclipsing probably is increased by 10\% in a
system with $e = 0.3$. On the other hand, the probability of two eclipses
occurring in a highly-eccentric orbit is reduced by around a half
\citep{2011ApJ...738..170M}. The available eccentricity-period distributions
have been obtained from the RV surveys. However, in order not to insert any
potential spectroscopic biases, we have adopted the zero eccentricity case.

\subsection{WASP sampling probability}
\label{det_prob}

For systems which do eclipse, we need to determine the probability that
WASP will have sufficient observations in order to be able to detect these
eclipses. This is the WASP sampling probability ($p_{\rm sample}$) and is
independent from the eclipsing probability. The diurnal observing pattern of
WASP, together with weather interruptions, affects the ability to detect
eclipsing systems. The average observing season is around 120 days, but
individual light curves range from less than 50 days up to nearly 200 days.

\begin{figure}[h]
\includegraphics[width=\columnwidth]{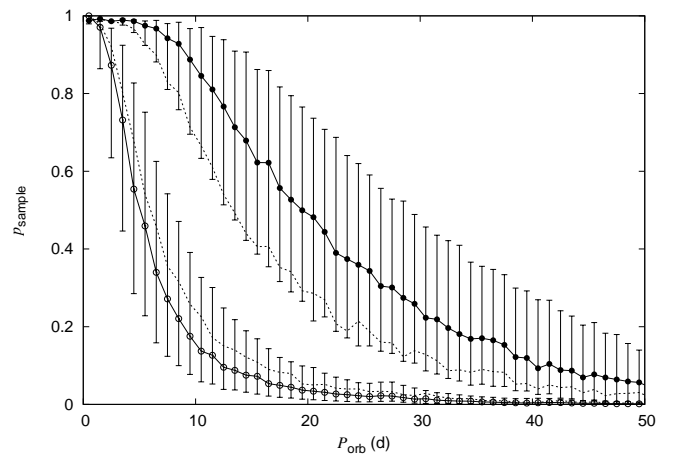}
\caption{The median expected WASP sampling probability ($p_{\rm sample}$) as a function of orbital
period ($P_{\rm orb}$). The filled circles are the case for a system of two identical stars,
while the open circles are that for a star with a small dark companion. The
dotted lines demonstrate the large change in sampling probabilities between the
two eclipses per orbit case and that for a single eclipse, calculated for a
$0.5\,R_{\sun}$ companion.}
\label{prob-det}
\end{figure}

In order to determine the expected WASP sampling probability, we require
a minimum of 3 eclipses within a single-season of WASP data and assume that at
least 10 data points within each eclipse are required for a detection. The
probabilities were obtained using a method similar to
\citet{2001PASP..113..439B}. Trial periods in the range 0.7 to 100 days in
0.02-day steps were used to determine the fractional phase detection probability
at each period. The individual probabilities were calculated for all the
observations of the 1742 Am stars using the actual time sampling and combined to
give the median sampling probability as a function of orbital period.
Again, we considered the two cases, small dark companion and two equal stars. In
the latter the sampling probability is significantly increased due to the
presence of two eclipses per orbital period, where \textsc{hunter} would
preferentially detect the period as half that of the true period. The
probability distribution is smoothed by binning into 1-day period bins
(Fig.\,\ref{prob-det}). The sampling probability drops as the size of the
companion decreases and as the period increases. The uncertainty, as given by
the lower and upper quartile values, is as expected quite large and is the
dominant source of uncertainty in the overall detection probability. The
transition between the two detectable eclipses per orbit and the single case
occurs around $R_2 = 0.5\,R_{\sun}$, which is approximately late type-K
spectral type.

\subsection{Dilution due to blending}

The relatively large pixel size of WASP data makes it susceptible to blending by
other stars within the photometry aperture. This dilution will mean that the
detection of shallow eclipses will be less efficient. Thus, WASP data might
systematically under estimate the number of such systems. Using the NOMAD $r$
magnitudes \citep{2004AAS...205.4815Z}, we have determined the amount of blending expected within the
48{\arcsec} WASP photometry aperture for our sample of Am stars. Around 48\%
have no blending, and 80\% have a dilution of $<$0.1. Fewer than 3\% of the
sample have dilution $>$0.5. For the case when dilution is 0.5, the minimum
actual eclipse depth would be 0.02\,{mag.}, corresponding to the observed
0.01\,{mag.} limit as above. Hence, $\delta R$ would become 0.28 and 0.23, for
a 0.15\,$R_1$ dark companion and two equal stars, respectively, compared
to 0.17 and 0.15 for the undiluted case. Hence, not only is the probability of
detecting an eclipse reduced by around 10\%, but also the lower radius limit is
increased.

On the other hand, blending also raises the possibility that any detected
eclipse is actually on a nearby fainter star within the WASP aperture. For
example, Renson\,28390 (HD\,98575A) is an 8.9\,{mag.} Am star and was
originally selected as a binary system with $\sim$0.01\,{mag.} eclipses
on an 1.5778\,d period. However, targeted follow-up photometry using TRAPPIST
\citep{2011Msngr.145....2J} revealed that the eclipse is actually on the
12.5\,{mag.} star situated 16{\arcsec} away. Thus, some of the eclipses reported
here may not be on the Am star. Only by targeted photometry can we be absolutely
sure.

\subsection{Overall probability}
\label{overall_prob}

\begin{figure}[h]
\includegraphics[width=\columnwidth]{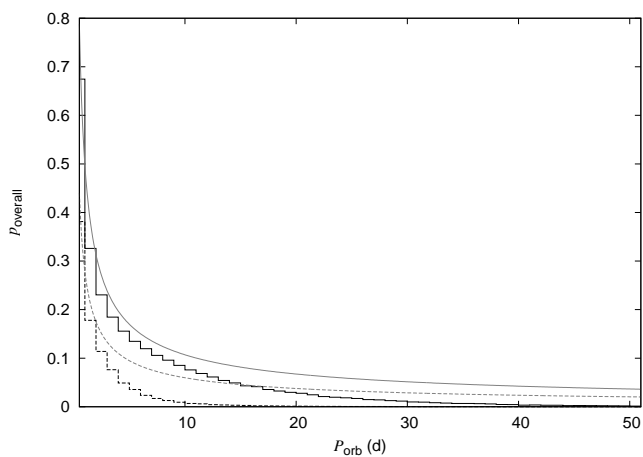}
\caption{The overall expected WASP detection probability ($p_{\rm
overall}$)
as function of orbital period ($P_{\rm orb}$).
Solid histogram is the case for a system with two identical stars,
while the dashed histogram is that for a star with a small dark companion.
The corresponding eclipsing probabilities discussed in Sect.\,\ref{ecl_prob}
are given as grey lines for reference.}
\label{prob-overall}
\end{figure}

The overall probability of finding binary systems with WASP data ($p_{\rm
overall}$) is the product of the eclipsing and sampling probabilities
(Fig.\,\ref{prob-overall}). Dilution is not significant in most of the stars
surveyed, so will be neglected. The overall detection probability for the small
dark companion case is in agreement with that obtained by
\cite{2012A&A...548A..48E} in their evaluation of the planetary transit
detection performance of WASP data using Monte-Carlo simulations. It is
worth remembering that these probabilities have a relatively large uncertainty,
especially the sampling probability. Nevertheless, these will enable us to
explore the population of Am binary systems.

\section{Discussion}

\subsection{Expected period distribution of Am binaries}

\begin{figure}[h]
\includegraphics[width=\columnwidth]{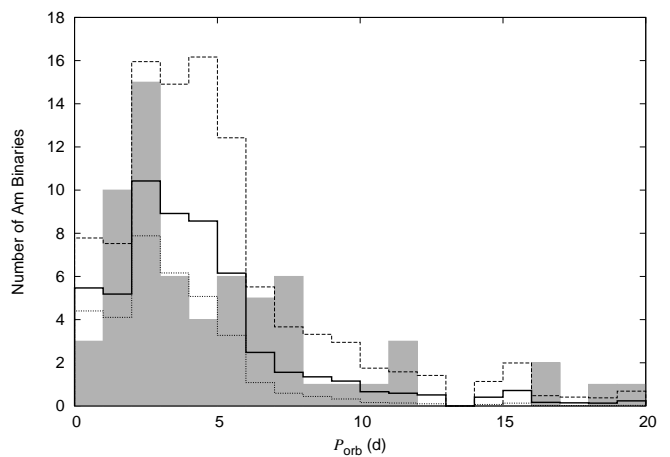}
\caption{Period distribution of eclipsing Am star binaries. The WASP
eclipsing Am star distribution is given as solid grey. The predicted period
distribution based on the results of spectroscopic binaries are given as
dashed-line for the two identical stars and dotted-line for the dark companion
case. The thicker solid-line is that predicted based on the ratio of SB1 and SB2
systems.}
\label{dist-predicted}
\end{figure}

The results from the radial velocity studies of \cite{1985ApJS...59..229A} and
\cite{2007MNRAS.380.1064C} can be used to predict the number and period
distribution of eclipsing Am binaries. The combined sample comprises 151 Am
stars, with 61 SB1s and 28 SB2s. This was binned onto 1-day bins, normalized by
the total number of stars, to generate a period probability distribution for Am
stars. Multiplying by the estimated WASP detection probability ($p_{\rm
overall}$) obtained in Sect.\,\ref{DetProb} and by the number of stars
in the WASP sample (1742), yields an estimate of the expected period
distribution of eclipsing Am stars. In Figure\,\ref{dist-predicted} the
WASP eclipsing Am star period distribution is compared to that predicted for the two identical stars
and the dark companion cases. Since these represent the extrema of the
probabilities, we also include the predicted distribution obtained using the
ratio of SB1 and SB2 systems from the RV studies.

The number of eclipsing Am stars found by WASP does appear to be broadly
consistent with the expected number of systems. We recall from
Sect.\,\ref{introduction} that the fraction of spectroscopic binaries is
60$\sim$70\%. Thus, the eclipsing fraction appears to be similar, suggesting a
significant fraction of Am stars might be single or have hard to detect
companions. The period distribution is, however, slightly different, with a
pronounced peak at shorter periods due to the inclusion of close binaries. The
$v \sin i$ distributions of both \cite{1985ApJS...59..229A} and
\cite{2007MNRAS.380.1064C} are skewed toward lower values than the
\citet{2009A&A...498..961R} sample. These radial velocity studies have
preferentially avoided stars with high rotation, which accounts for the excess
of short period systems found in the WASP sample. The distribution of
Am-type spectroscopic binaries in the \citet{2009A&A...498..961R}
catalogue (some 210 systems) shows a similar short orbital period excess,
due to the inclusion of Am stars with a wide range of rotational
velocities. 

\subsection{Mass ratio distribution}

\begin{table*}
\caption{Results from \textsc{jktebop} fits to light curves (Detached
systems only) and approximate stellar parameters, assuming primary is
${T_\mathrm{eff}}_A = 7500$\,K. See Text for details.}
\label{JKTEBOP}
\begin{tabular}{llllllllllll}\hline\hline
Renson &$\frac{R_A+R_B}{a}$ & $R_B/R_A$ & $i$ & $e$ & $J_B/J_A$ & $P$ & ${T_\mathrm{eff}}_B$ & $R_A$ & $R_B$ & $q$ & $q$ (literature) \\ \hline
 3750& 0.307& 0.111& 89.95&  0.00&  0.00&  1.2199& 2000\S&  1.63&  0.18&  0.07 &   $<$0.003    {\citep{2010MNRAS.407..507C}}\\
 4660& 0.298& 0.910& 87.61&  0.00&  0.84&  2.7807& 7176&  1.97&  1.80&  0.92 &   0.91        {\citep{2011MNRAS.414.3740S}}\\
 6720& 0.344& 1.354& 79.59&  0.00&  0.91&  2.3832& 7333&  1.68&  2.26&  1.07 &\\
 7310& 0.161& 0.601& 86.37&  0.00&  0.50&  6.6637& 6322&  2.20&  1.32&  0.71 &\\
 7730& 0.156& 3.000& 89.57&  0.24&  0.10& 39.2827& 4218&  2.93&  8.71&  0.84 &   0.96        {\citep{1988AJ.....96.1040P}}\\
 8215& 0.105& 0.940& 86.43&  0.00&  0.44&  8.7959& 6126&  1.38&  1.29&  0.77&\\
 9237& 0.248& 1.036& 80.52&  0.00&  0.49&  2.8562& 6288&  1.51&  1.55&  0.81&\\
 9318& 0.148& 0.472& 85.89&  0.06&  0.72& 11.1133& 6901&  3.24&  1.52&  0.71&\\
 9410& 0.189& 0.634& 87.74&  0.02&  0.30& 16.7873& 5537&  5.07&  3.19&  0.59&   0.81{\dag}  {\citep{1971AJ.....76..544L}}\\
10016& 0.242& 1.323& 86.98&  0.25&  1.03&  5.4309& 7557&  2.11&  2.81&  1.11&\\
11100& 0.165& 0.397& 85.07&  0.00&  0.36&  4.0375& 5818&  1.79&  0.71&  0.61&\\
11470& 0.228& 0.580& 89.36&  0.00&  0.39&  2.8721& 5931&  1.75&  1.02&  0.66&\\
14850& 0.215& 0.272& 83.02&  0.00&  0.08&  3.9795& 3925&  2.43&  0.66&  0.30&\\
15034& 0.202& 0.469& 83.88&  0.00&  0.54&  7.5392& 6423&  3.39&  1.59&  0.65&\\
15190& 0.174& 0.439& 82.60&  0.00&  0.02&  5.1229& 2952&  1.97&  0.86&  0.21&   0.33{\dag}  {\citep{2003MNRAS.346..555C}}\\
15445& 0.270& 0.881& 86.32&  0.00&  0.92&  5.7606& 7355&  3.10&  2.72&  0.93&\\
21400& 0.106& 0.513& 86.63&  0.00&  0.04&  7.7729& 3435&  1.49&  0.77&  0.29&\\
22860& 0.136& 0.558& 83.84&  0.28&  2.81&  7.9006& 9712&  2.33&  1.30&  1.17&\\
25020& 0.188& 0.081& 89.10&  0.00&  0.00&  6.6284& 2000\S&  3.39&  0.28&  0.06&\\
25070& 0.185& 0.500& 85.60&  0.14&  0.91&  6.7149& 7323&  2.85&  1.42&  0.79&   1.00 (this work)\\
29290& 0.314& 0.977& 79.45&  0.00&  0.95&  3.2922& 7406&  2.30&  2.23&  0.98&\\
30090& 0.221& 0.788& 82.26&  0.00&  0.97&  4.3486& 7451&  2.11&  1.67&  0.93&\\
30110& 0.148& 0.162& 84.55&  0.00&  0.09&  4.3145& 4108&  1.92&  0.31&  0.37&\\
30457& 0.137& 0.231& 86.82&  0.00&  0.00& 11.9420& 2000\S&  3.20&  0.74&  0.08&\\
30650& 0.060& 0.378& 88.38&  0.00&  0.00& 18.1210& 2000\S&  1.53&  0.58&  0.09&\\
35000& 0.190& 0.835& 82.07&  0.00&  0.00& 10.2861& 2000\S&  2.69&  2.26&  0.13&\\
36660& 0.096& 0.698& 88.61&  0.41&  0.71& 16.3653& 6892&  2.30&  1.60&  0.81&   0.80 (this work)\\
37220& 0.135& 0.962& 85.57&  0.00&  0.97&  5.7931& 7437&  1.38&  1.33&  0.98&\\
37610& 0.240& 1.455& 85.50&  0.00&  1.28&  3.2387& 7974&  1.38&  2.02&  1.19&\\
38500& 0.173& 0.348& 83.65&  0.00&  0.19&  3.9938& 4945&  1.87&  0.65&  0.47&\\
40350& 0.156& 0.272& 88.22&  0.00&  0.00&  7.0687& 2000\S&  2.43&  0.66&  0.08&\\
40910& 0.132& 1.114& 87.17&  0.05&  1.05& 11.1163& 7594&  2.03&  2.23&  1.05&\\
42906& 0.062& 0.860& 87.84&  0.00&  0.00& 19.6987& 2000\S&  1.23&  1.06&  0.12&   0.98        {\citep{2003MNRAS.346..555C}}\\
44140& 0.309& 1.391& 86.51&  0.02&  0.66&  2.0598& 6763&  1.31&  1.81&  0.96&   0.90        {\citep{1970ApJ...162..925P}}\\
49380& 0.161& 0.355& 85.06&  0.11&  0.57&  5.0204& 6508&  2.14&  0.76&  0.68&   0.83 (this work)\\
51506& 0.322& 0.119& 80.30&  0.12&  0.17&  1.9196& 4787&  2.69&  0.32&  0.46&   0.12 (this work)\\
56310& 0.285& 1.108& 84.26&  0.00&  0.88&  2.6959& 7258&  1.68&  1.85&  0.99&\\
58170& 0.366& 0.854& 88.24&  0.00&  0.48&  1.6047& 6246&  1.65&  1.42&  0.77&   0.78        {\citep{1968ApJ...154..191P}}\\
58256& 0.271& 0.378& 81.63&  0.00&  0.41&  2.9673& 5983&  2.50&  0.95&  0.59&\\
59780& 0.228& 1.200& 87.87&  0.33&  0.95&  7.3514& 7412&  2.61&  3.15&  1.06&   0.98        {\citep{1999AJ....118.1831T}}\\
\hline
\end{tabular}
\tablefoot{\dag Mass ratio (q) obtained using spectroscopic binary mass
function, $f(m)$, and assuming $M_1$ = 1.7\,$M_\sun$ and $i$ = 90.\newline
\S lower-limit on {${T_\mathrm{eff}}_{B}$} imposed when $J_B/J_A = 0$.
}
\end{table*}

Without direct determinations of masses from spectroscopic studies, we can only
make a rather crude estimate of the mass distribution of the eclipsing
systems from their light curves and the \textsc{jktebop} fits. Since the
bolometric correction for late-A stars is small, we can make the approximation
that the ratio of bolometric surface brightnesses is given by the WASP bandpass surface brightness ratio
($J_B/J_A$). Thus the effective temperature of the secondary
(${T_\mathrm{eff}}_{B}$) can be obtained from,
\[{T_\mathrm{eff}}_{B} \approx {T_\mathrm{eff}}_{A} \times (J_B/J_A)^{1/4},\]
where the effective temperature of the primary (${T_\mathrm{eff}}_{A}$) is
assumed to be 7500\,K. With initial mass estimates of $M_A$ = $M_B$ = 1.7, the
known orbital period ($P_{\rm orb}$) and sum of the radii ($\frac{R_A+R_B}{a}$)
from \textsc{jktebop}, we determine initial values for $R_A + R_B$. Using the
ratio of the radii also from \textsc{jktebop} ($R_B/R_A$) we can determine
individual values for $R_A$ and $R_B$. Next, the \cite{2010A&ARv..18...67T}
relations are used to determine stellar mass estimates, by varying $\log g$ so
that radii obtained from \cite{2010A&ARv..18...67T} agrees with that expected
from the \textsc{jktebop} analysis. The procedure is iterated until there are no
changes in parameters. The results are presented in Table\,\ref{JKTEBOP}, along
with values for mass ratio ($q$) from spectroscopic analyses in the literature
and those determined in Section\,\ref{Data_modelling}. The average difference
between our estimated $q$ values and the spectroscopic values is $-$0.11,
but with an rms scatter of $\pm$0.25.

\begin{figure}[h]
\includegraphics[width=\columnwidth]{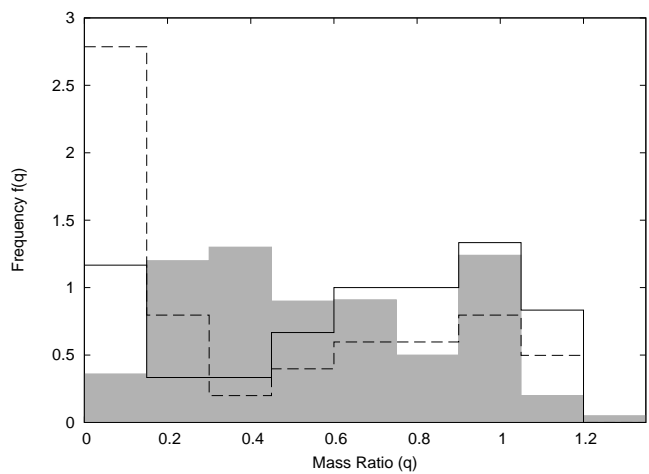}

\caption{Mass ratio distribution for Am binary systems. The solid-line
histogram is the distribution based on eclipsing binaries in the current work,
while the solid grey histogram is that presented by
\cite{2010A&A...524A..14B} based on the spectroscopic sample. The dashed
histogram is the estimated eclipsing binary distribution after making allowance
for the WASP detection probabilities.}

\label{q-dist}
\end{figure}

As discussed in Sect.~\ref{ecl_prob}, eccentric systems may not always show two eclipses when both stars are similar.
For example Renson\,42906 (HD\,151604; V916 Her) is an eccentric ($e = 0.566$)
system with mass ratio close to unity \citep{2007MNRAS.380.1064C}, but the WASP
data only shows one eclipse per orbit and value of $q=0.12$. Thus, these systems
will appear to have anomalously low $q$ values, further adding to the
uncertainty in the mass-ratio distribution.

\cite{2010A&A...524A..14B} concluded that the mass-ratio distribution showed
hints of a double-peaked distribution, with peaks at $q \sim 0.3$ and $q \sim
1$. The mass-ratio distribution based on the estimated properties from the WASP
light curves for detached systems is noticeably different (Fig.\,\ref{q-dist}).
The estimated WASP mass-ratio distribution shows a broad peak near unity,
with a deficit around $q \simeq 0.3$. However, the WASP detection probability
varies with companion size (Sect.\,\ref{det_prob}). Assuming that companions
with masses around 0.5\,{$M_\sun$} correspond to the transition between the two
system scenarios discussed earlier, there would be an increase in the number of
low $q$ systems relative to the high $q$ systems. The distribution would
become flatter,  similar to that found by \cite{2010A&A...524A..14B}, who noted
that a flat mass-ratio distribution also appeared to be a good fit. While
there are genuinely low $q$ systems (e.g. Renson 3750), the apparent excess of
such systems may not be real, since some of these may be pairs of similar stars
with a true period equal to twice the assumed one (as noted with a dagger in
Table~\ref{table:Am Binaries}). Hence, since our mass-ratio estimates are based
on photometry alone, radial velocity studies are required to determine
spectroscopic mass-ratios of the whole sample, before any firmer conclusions can
be drawn.

\section{Summary}

A survey of 1742 Am stars using light curves from the SuperWASP project has found
70 eclipsing systems, of which 28 are previously unreported detections and 4 are
suspected eclipsing systems. While this represents only 4\% of the sample, after
correction for eclipsing and WASP detection probabilities, the results are
consistent with 60--70\% incidence of spectroscopic binaries found from radial
velocity studies \citep{1985ApJS...59..229A,2007MNRAS.380.1064C}. This indicates
that there is not a deficit of eclipsing Am binary systems, as suggested by 
\cite{1990clst.book.....J}.

Like the radial velocity studies, the WASP study suggests that around 30--40\%
of Am stars are either single or in very wide systems. The WASP survey is able
to detect low-mass stellar and sub-stellar companions that were below the radial
velocity studies' detection limits. Thus, systems like HD\,15082 (WASP-33) would
not form part of the spectroscopic mass distribution. On the other hand, the
WASP survey is unable to detect compact companions, such as white dwarfs, which
would, if present, have been detected in the radial velocity studies. The
average mass of a white dwarf is around 0.6\,{$M_\sun$}
\citep{2013ApJS..204....5K}, corresponding to a $q$ of around 0.3$\sim$0.4. The
only short-period system with a white dwarf companion in the
\citet{2009A&A...498..961R} catalogue is HD\,204188 (IK Peg)
\citep{1993MNRAS.262..277W}, suggesting that such objects are relatively rare
\citep{2013MNRAS.435.2077H}.

Using \textsc{jktebop} fits to the WASP light curves, estimates of mass-ratios
have been determined. The WASP mass-ratio distribution is consistent with that
obtained from the spectroscopic studies \citep{2010A&A...524A..14B}. However, if
an approximate allowance is made for WASP detection probabilities there is a
suggestion of an excess of low mass-ratio systems. While this could be
explained by the presence of sub-stellar companions to Am stars, it is more
likely that this is due to pairs of similar stars with true periods twice that
assumed or the presence of eccentric systems exhibiting only one eclipse.
Hence, radial velocity studies of the eclipsing systems found with WASP are
required in order to fully explore the mass-ratio distribution of Am binary
systems.

\begin{acknowledgements}

The WASP project is funded and operated by Queen's University Belfast, the
Universities of Keele, St. Andrews and Leicester, the Open University, the Isaac
Newton Group, the Instituto de Astrofisica de Canarias, the South African
Astronomical Observatory and by STFC. OIP's contribution to this paper was
partially supported by PIP0348 by CONICET. MG and EJ are Research Associates of
the Belgian Fonds National de la Recherche Scientifique (FNRS). LD is FRIA PhD
student of the FNRS. TRAPPIST is a project funded by the FNRS under grant FRFC
2.5.594.09.F, with the participation of the Swiss  National Science Foundation
(SNF). This research has made use of the SIMBAD database, operated at CDS,
Strasbourg, France. We thank the anonymous referee for his thoughtful
and constructive comments on the original manuscript.

\end{acknowledgements}


\end{document}